\documentclass[reprint,aip,superscriptaddress,amsmath,amssymb,floatfix]{revtex4-1}
\usepackage{graphicx}
\usepackage{layout}
\usepackage{dcolumn}
\usepackage{braket}
\usepackage{bm}
\usepackage{color}
\usepackage{verbatim}
\usepackage{txfonts}
\usepackage{gensymb}

\usepackage{natbib}
\begin{document}

\title{Patterning Superconductivity in a Topological Insulator}

\author{Jerome T. Mlack}
\affiliation{Department of Physics and Astronomy, Johns Hopkins University, Baltimore, Maryland 21218, USA}
\affiliation{Department of Physics and Astronomy, University of Pennsylvania, Philadelphia, Pennsylvania 19104, USA}

\author{Atikur Rahman}
\affiliation{Department of Physics and Astronomy, Johns Hopkins University, Baltimore, Maryland 21218, USA}

\author{Gopinath Danda}
\affiliation{Department of Physics and Astronomy, University of Pennsylvania, Philadelphia, Pennsylvania 19104, USA}
\affiliation{Department of Electrical and Systems Engineering, University of Pennsylvania, Philadelphia, Pennsylvania 19104, USA}

\author{Natalia Drichko}
\affiliation{Department of Physics and Astronomy, Johns Hopkins University, Baltimore, Maryland 21218, USA}

\author{Sarah Friedensen}
\affiliation{Department of Physics and Astronomy, University of Pennsylvania, Philadelphia, Pennsylvania 19104, USA}

\author{Marija Drndic}
\affiliation{Department of Physics and Astronomy, University of Pennsylvania, Philadelphia, Pennsylvania 19104, USA}

\author{Nina Markovic}
\affiliation{Department of Physics and Astronomy, Johns Hopkins University, Baltimore, Maryland 21218, USA}
\affiliation{Department of Physics and Astronomy, Goucher College, Baltimore, Maryland 21204, USA}


\maketitle

\textbf{Topological superconductors are predicted to host Majorana fermions\cite{Fu_Kane_2008,Beenakker_Review}, which may be used as building blocks for fault-tolerant quantum computing\cite{Kitaev, Nayak}. While some evidence of topological superconductivity has been found in doped bulk topological insulators\cite{Hor_2010,Hor_2011,TV_Bay,M_Kreiner,S_Satoshi,ZLiu,Matano} and Majorana fermions have been reported in one-dimensional systems in proximity to superconductors \cite{Mourik,Das,Albrecht,Nadj}, one of the remaining challenges is to find a convenient experimental platform for realizing circuits that would allow a pair-wise exchange of Majorana fermions known as braiding\cite{Nayak}. Here we show that superconductivity can be patterned directly into a topological insulator Bi$_2$Se$_3$ by doping selected regions with palladium, using electron beam lithography and in-situ annealing. Electrical transport measurements at low temperatures show superconducting transitions in the doped regions, while structural characterization techniques indicate that Pd remains localized in targeted areas. Our results show that it is possible to pattern superconducting circuits of arbitrary shapes in a topological material, offering a promising way for building topologically protected quantum devices.}

Bi$_2$Se$_3$ is a widely studied topological insulator \cite{Hasan_Kane_Rev2010} that is known to become superconducting upon doping with copper or other metalllic elements\cite{Hor_2010, M_Kreiner, ZLiu}. The bulk Cu-doped Bi$_2$Se$_3$ was predicted to have topological properties \cite{Fu_Berg}, which have been reported in some experiments \cite{S_Satoshi, ZLiu, Matano}, while others indicated conventional superconductivity in these materials \cite{N_Levy,H_Peng}. It has been shown that the stability and the properties of Cu-doped Bi$_2$Se$_3$ depend critically on growth and quenching conditions\cite{Schneeloch}. An alternative to bulk doping is to induce superconductivity on the surface of Bi$_{2}$Se$_{3}$ by proximity to a conventional superconductor \cite{G_Koren_2011, BSacepe, JR_Williams_2012, S_Cho}. While both the bulk-doped and proximity-coupled systems are convenient platforms for studying topological superconductivity, harnessing their topological properties remains to be a challenge due to the difficult device architecture and poor electrical contacts. 
 
Some of these issues could be effectively resolved by combining the above approaches in order to create native superconducting and normal regions in desired patterns within the crystal of a topological insulator. By providing superconductivity in desired locations with clean interfaces, this will lend itself to patterning devices, which should enable further progress in the study of topological superconductivity.

\begin{figure*}[ht]
\centering
\includegraphics{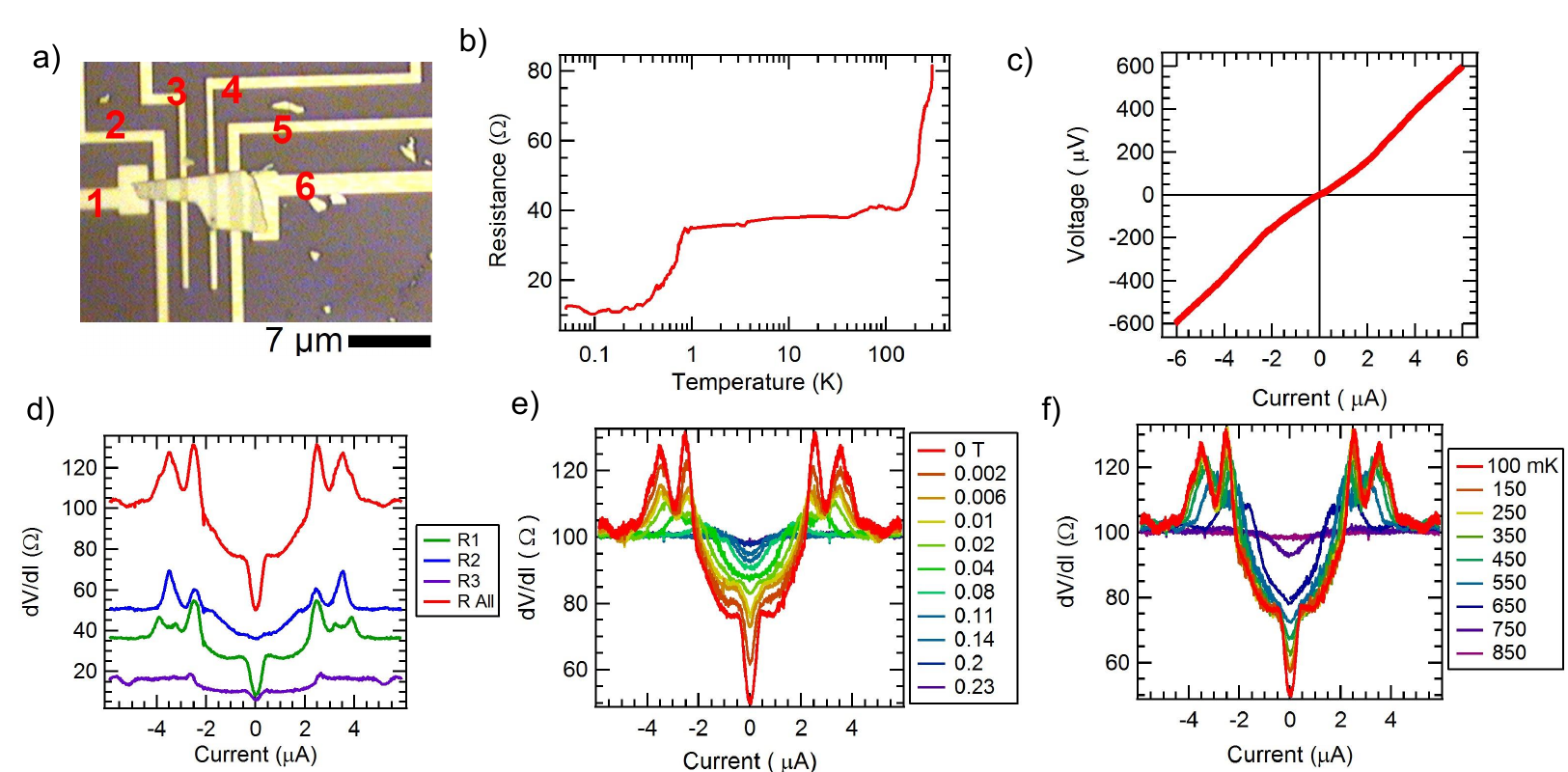}
\caption{\label{target_combined}
Electrical measurements of a Pd-doped Bi$_{2}$Se$_{3}$ device after annealing. (a) Optical microscope image of a sample after annealing under flowing Ar gas at 220 $^\circ$C for one hour. The irregular shape in the center is an exfoliated Bi$_{2}$Se$_{3}$ flake, and the Pd leads are marked by numbers 1-6. The parts of the Bi$_{2}$Se$_{3}$ flake that had been covered with 90nm of Pd appear darker in the image. (b) Four-probe resistance as a function of temperature with current sourced across leads 1 and 6 and voltage measured across leads 2 and 3. A transition to a lower resistance state is observed around 800 mK.(c) Four-probe V-I curve with current sourced from leads 1 to 6 and voltage measured across leads 2 and 5. (d) Four-probe differential resistance measurements with current sourced from lead 1 to 6 and voltage measured across leads 2 and 5 (red), leads 2 and 3 (green, Region 1), leads 3 and 4 (blue, Region 2), and leads 4 and 5(purple, Region 3). All three regions show transitions, but at slightly different bias currents. Differential resistance measurements of the whole device, with voltage measured across 2 and 5, is shown as a function of magnetic field (e) and temperature(f). In both sets of data, two primary peaks are observed at 2.5 $\mu$A and 3.5 $\mu$A. 
}
\end{figure*}

 Bi$_{2}$Se$_{3}$ nanocrystals were mechanically exfoliated onto 300 nm SiO$_2$ on Si substrates. Palladium leads were fabricated onto the nanocrystals using standard electron beam and optical lithography methods. The devices were annealed in a quartz tube furnace at a set temperature between 200 and 300 $^\circ$C while flowing argon gas through the tube at a rate of 200 sccm. The set temperature was maintained for one hour, after which the tube furnace was turned off and allowed to cool down naturally. An optical image of a device that was annealed at 220 $^\circ$C is shown in Fig.\ref{target_combined}(a). A second layer of palladium leads was then added over the original leads in order to ensure good contact for electrical measurements. 

A four probe measurement of the resistance as a function of temperature between leads 2 and 3 of Fig.\ref{target_combined}(a) is shown in Fig.\ref{target_combined}(b). A transition to a low-resistance state is observed at 800 mK. This region is not fully superconducting and the resistance plateaus at approximately 12$\Omega$. The current-voltage measurement across the entire device, with current sourced between leads 1 and 6 and voltage measured across leads 2 and 5, is shown in Fig.\ref{target_combined}(c). The non-linear behavior above 2 $\mu$A  is emphasized in the calculated differential conductance, measured on different regions of the device, shown in Fig.\ref{target_combined}(d). 

All three measured regions show peaks in the dV/dI curve at 2.6 $\mu$A as well as two valleys, with the central valley spanning $\pm$ 2.6 $\mu$A and the wider valley spanning $\pm$ 5 $\mu$A. Region 1 shows secondary and tertiary peaks at 3.2 $\mu$A, and 3.7 $\mu$A. Region 2 shows a secondary peak at 3.4 $\mu$A. Region 3 exhibits the primary peak at 2.6 $\mu$A and a secondary peak at  5.8 $\mu$A. Temperature and magnetic field dependence of the differential resistance across the entire device is shown in Fig.\ref{target_combined}(e) and (f), respectively. In both sets of data two primary peaks are observed at 2.5 $\mu$A and 3.5 $\mu$A. As either the temperature or magnetic field is increased, the dip in resistivity at low currents gradually disappears. A similar behavior of dV/dI has been reported previously in Bi$_{2}$Se$_{3}$ samples in which superconductivity was induced on the surface by proximity to a conventional superconductor \cite{JR_Williams_2012, BSacepe}.

\begin{figure*}[ht]
\centering
\includegraphics{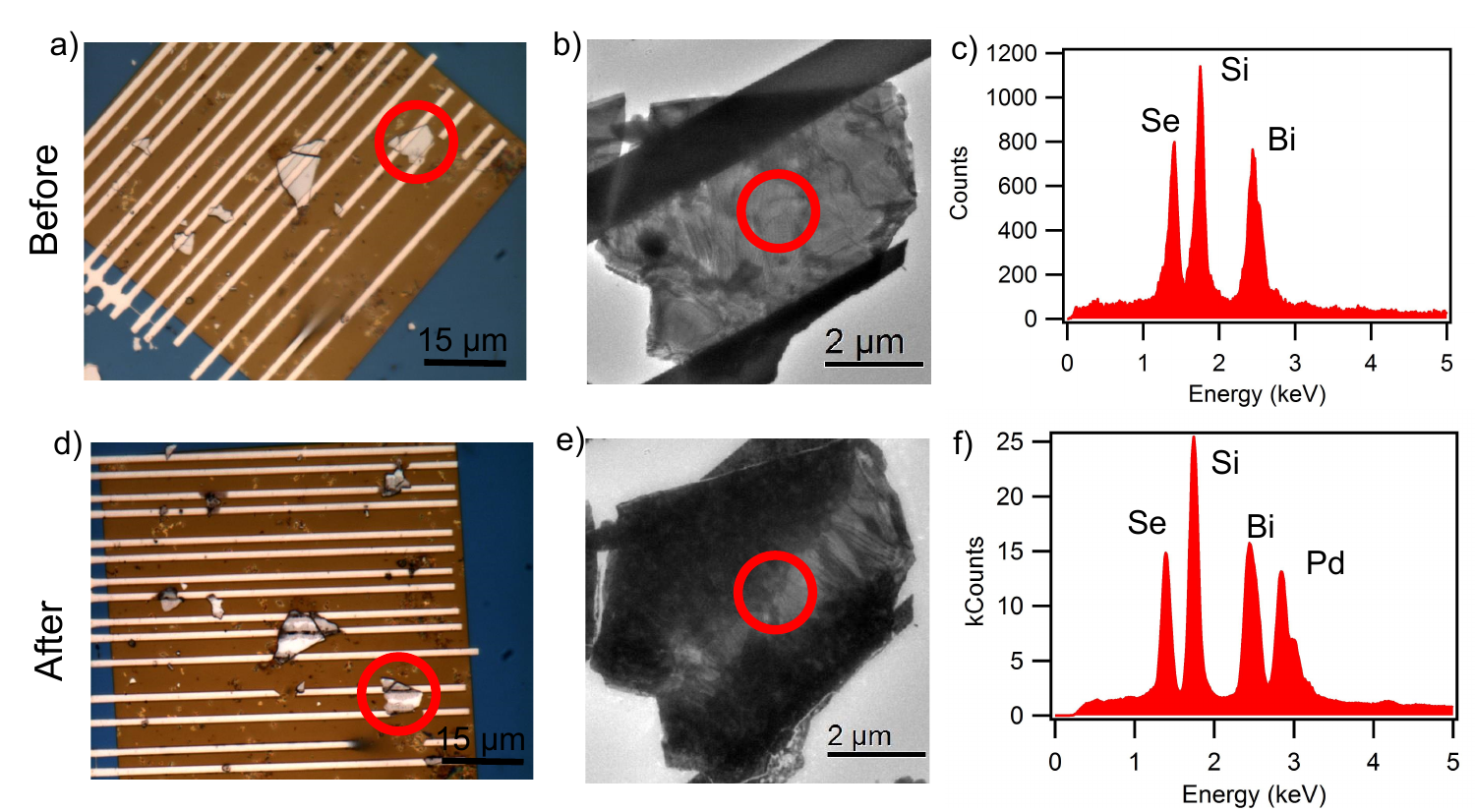}
\caption{\label{TEM_BA}
Images and EDS spectra of exfoliated Bi$_{2}$Se$_{3}$ flakes before and after thermal annealing. (a) Optical image of Bi$_{2}$Se$_{3}$ flakes exfoliated onto a silicon nitride window and covered with lines of palladium. (b) TEM image of flake with palladium line, circled in red in (a). (c) EDS spectra of red circled area in (b) showing composition of Bi$_{2}$Se$_{3}$ and signal from the silicon nitride window. (d) Optical image of flakes shown in (a) after thermal annealing for 1 hour at 295 $^\circ$C under argon gas flowing at 200 sccm. (e) TEM image of flake, circled in red in (d), post-annealing. (f) EDS spectra, from location circled in red in (e), post annealing, indicating that palladium has entered into the flake.
}
\end{figure*}

\begin{figure*}[ht]
\centering
\includegraphics{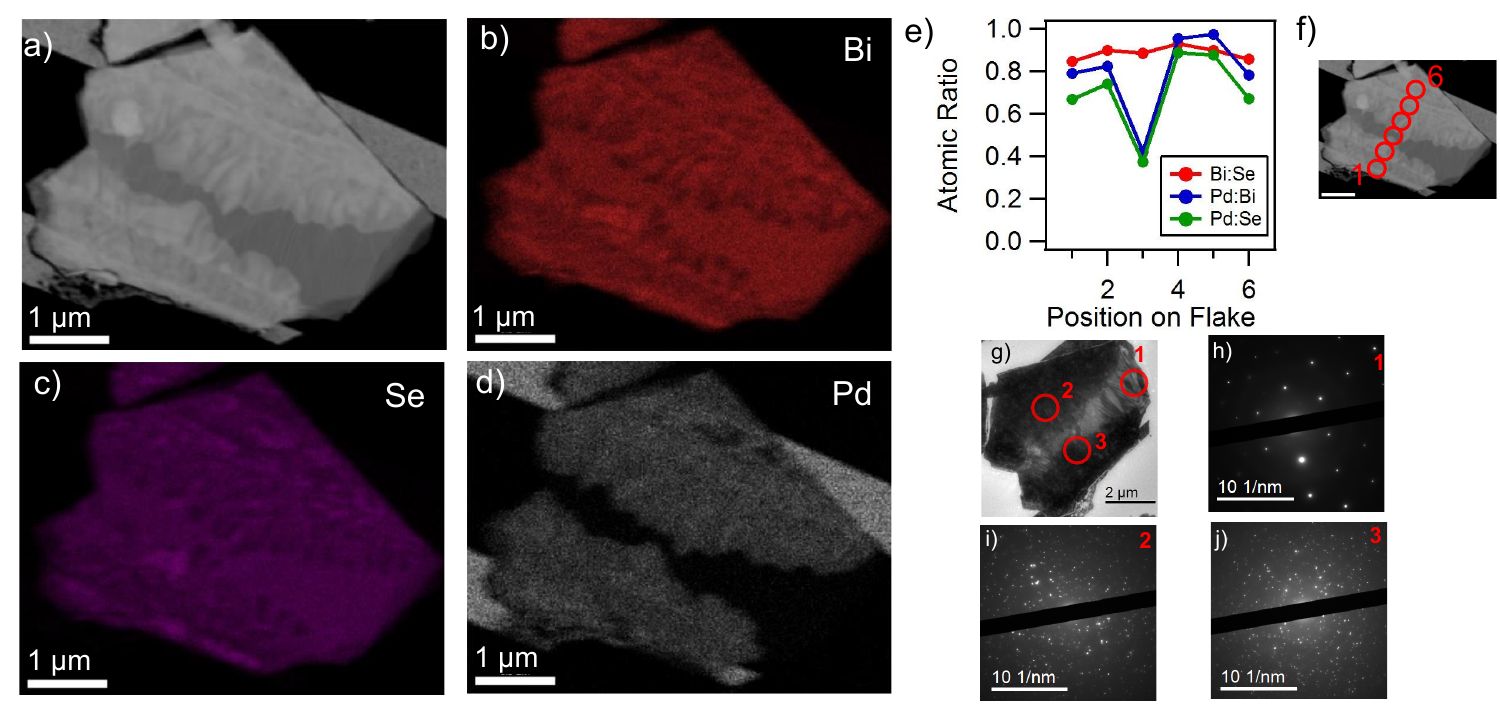}
\caption{\label{EDS_Map}
Elemental EDS maps of an annealed Bi$_{2}$Se$_{3}$ flake depicted in Fig.\ref{TEM_BA}(e), atomic ratios across the cross-section of the flake, and selected area diffraction images from several positions on the flake. (a) Counts per second map of the annealed flake. (b) Mapping of bismuth. (c) Mapping of selenium. (d) Mapping of palladium. (e) Atomic ratios of Bi:Se(red), Pd:Bi(blue), and Pd:Se(green) at several locations along the cross section of the flake as depicted in (f), starting with position 1 in the lower left hand corner to position 6 in the upper right. (g) Image of flake with regions chosen for diffraction identified. (h) SAED from Region 1 in (g), showing a hexagonal crystal lattice with radial pattern corresponding to [11$\bar{2}$0] and lattice constant a = 0.413nm. (i) SAED from Region 2 in (g), showing a polycrystalline pattern. (j) SAED from Region 3 in (g), showing a polycrystalline pattern.
}
\end{figure*}

In order to investigate what happens to the Pd during the annealing process, samples were imaged using a transmission electron microscope (TEM) before and after annealing, as shown in Fig. \ref{TEM_BA} and Fig. \ref{EDS_Map}. Bi$_{2}$Se$_{3}$ flakes were exfoliated onto a 100 nm thick silicon nitride TEM window with 90 nm thick palladium lines deposited across the flakes. Fig.\ref{TEM_BA}(a) shows an optical microscope image of a sample before annealing. A representative flake, circled in red in Fig.\ref{TEM_BA}(a), is shown in the TEM image in Fig.\ref{TEM_BA}(b). Fig.\ref{TEM_BA}(c) is the energy-dispersive X-ray (EDS) spectrum for the area of the flake circled in red in Fig.\ref{TEM_BA}(b). The EDS spectrum indicates that the central region of the flake is pure Bi$_{2}$Se$_{3}$ with a background Si signal from the substrate. 

The same sample is shown in Fig.\ref{TEM_BA}(d-f) after annealing at 290 $^\circ$C. In the optical image of Fig.\ref{TEM_BA}(d), the exfoliated flakes that were in contact with Pd show a clear change in color in the vicinity of the Pd lines. The TEM image in Fig.\ref{TEM_BA}(e) shows that the flake from Fig.\ref{TEM_BA}(b) has undergone a drastic change after annealing. The flake is overall less transparent and the contact to the Pd lines has disappeared, indicating that the Bi$_{2}$Se$_{3}$ flake has absorbed the palladium. Further evidence that the Pd has spread into the flake is provided by the EDS spectrum from the same region as Fig.\ref{TEM_BA}(c), which now includes a peak indicating the presence of Pd along with the original Bi, Se, and Si peaks.

In order to better understand the dynamics and extent of Pd migration in the Bi$_{2}$Se$_{3}$ flake, EDS map of the flake in Fig.\ref{TEM_BA} was acquired after annealing, as shown in Fig.\ref{EDS_Map}. It was observed in elemental spatial maps that the while the Bi (Fig.\ref{EDS_Map}b) and Se (Fig.\ref{EDS_Map}c) peaks were present across the entire flake, the Pd (Fig.\ref{EDS_Map}d) extended only up to a certain length into the flake from the leads, leaving the central region unaltered Bi$_{2}$Se$_{3}$. The weaker peak intensity of Pd at the lead-flake-substrate interface compared to lead-substrate interface and the uniform atomic ratio of Pd inside the flake, suggest the movement of Pd atoms from the leads to the flakes during annealing until a steady state concentration is reached.

In Fig.\ref{EDS_Map}(e), the atomic ratio of Bi:Se, Pd:Bi, and Pd:Se were calculated along a line profile (point scan area = 0.04 u$\mu$m$^2$) as shown in Fig.\ref{EDS_Map}(e). The ratio of Pd:Bi and Pd:Se were both seen to have a valley in the central region as expected due to the absence of Pd, consistent with the EDS Pd map. It was seen that the Bi:Se atomic ratio remained approximately constant throughout at a value slightly greater than 0.7 or 2:3. The fact that the Se concentration was slightly lower than expected indicates that some loss occurred as a result of the annealing process. While Bi$_2$Se$_3$ typically shows metallic properties in the bulk due to Se vacancies, these properties can be controlled by growth methods \cite{Brahlek}, chemical doping \cite{Kim}, or electrostatic doping\cite{Kim}. 

To further investigate the effect of Pd doping on Bi$_2$Se$_3$, electron diffraction results were obtained from three different regions, specified in Fig.\ref{EDS_Map}(g). Region 1 is an unaltered Bi$_2$Se$_3$ region, while regions 2 and 3 are from regions which were not covered by Pd before annealing. The diffraction in Region 1, shown in Fig.\ref{EDS_Map}(h), indicates the hexagonal Bi$_2$Se$_3$ radial pattern for the [11$\bar{2}$0] direction, with a calculated lattice parameter value of a = 0.413 nm. In the regions 2 and 3, which contain Pd, the diffraction data indicate a polycrystalline lattice structure, as shown in Figs.\ref{EDS_Map}(i) and (j) respectively. The constant Bi:Se atomic ratio and the presence of a polycrystalline diffraction pattern of Regions 2 and 3 could be suggesting that Pd is either altering or distorting the crystal structure without replacing the Bi or Se atoms. Another possibility is that there is a layer of polycrystalline Pd on top of the Bi$_2$Se$_3$. Further in-depth material characterization would be required for better understanding and tuning the Pd migration in Bi$_2$Se$_3$.

 Overall, the TEM analysis of the annealed flake shows that the Pd is absorbed by the Bi$_{2}$Se$_{3}$ flake during the annealing process and that the absorption occurs uniformly with a leading edge across the flake. Although superconductivity has been reported before in several compounds containing Pd and Bi or Se \cite{T_Takabatake_1988,T_Sakamoto_2008,R_Weihrich}, Bi:Se ratio suggests that it is unlikely that such compounds are forming in our samples. Since the Bi:Se ratio is fairly constant across the sample, the Pd is most likely intercalated in the Bi$_{2}$Se$_{3}$ flake. A substitution of Pd in Bi or Se sites would result in decrease in either Bi or Se counts in regions where Pd is present.

\begin{figure*}[ht]
\centering
\includegraphics{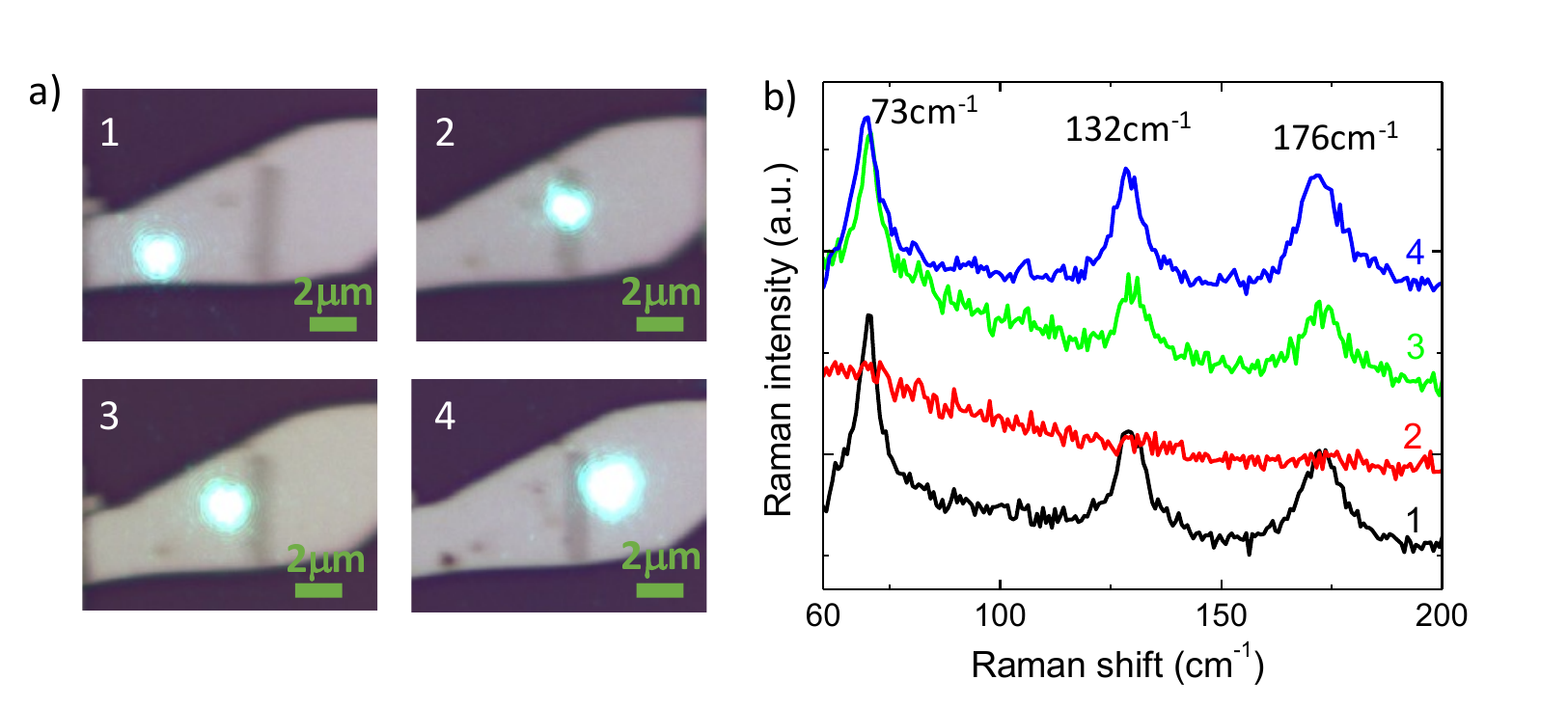}
\caption{\label{Raman}
Raman spectroscopy of Pd-doped Bi$_{2}$Se$_{3}$ flakes after annealing at 200 $^\circ$C(a) Optical images of a Bi$_{2}$Se$_{3}$ flake (light grey) exfoliated onto a silicon oxide substrate (black background). The darker grey vertical line in the center of the flake is a 90nm thick layer of Pd. The bright spot is the image of the Raman laser probe at positions 1(away from Pd), 2(on top of Pd), 3(directly to the left of the Pd) and 4 (directly to the right of the Pd). (b) Raman scattering spectra at positions 1-4 in the range of  Bi$_2$Se$_3$ phonons. Spectra from positions 1,3 and 4 show phonons of that are expected for pure Bi$_2$Se$_3$.  The spectrum from position 2 shows no bands, because the layer of Pd on top of Bi$_2$Se$_3$ is not transparent to light with the wavelength of 514 nm.
}
\end{figure*}

By comparing optical images and EDS data, we see that the regions penetrated by Pd appear gray on the optical images, providing a simple way to optically locate the Pd. Comparing samples annealed at different temperatures, we find that the extent of the Pd spreading can be controlled by the annealing temperature. At lower annealing temperatures, Pd is absorbed only in the targeted areas (as in Fig.\ref{target_combined} and Fig.\ref{Raman}), while at higher annealing temperatures, it also spreads away from the targeted areas (as in Fig.\ref{EDS_Map}). 
Additional measurements taken over several weeks indicate that the samples do not degrade with time, and the Pd remains in place.

Further evidence that the Bi$_{2}$Se$_{3}$ crystal structure remains intact upon annealing is provided by Raman spectroscopy. Raman scattering spectra were measured in backscattering geometry using Horiba Jobin-Yvon T64000 spectrometer equipped with Olympus microscope. 514.5 nm line of Ar$^+$ laser was used for excitation. The laser power was kept below 1 mW to avoid overheating of the sample. The size of the laser probe was about 2 $\mu$m in diameter. The spectra were recorded with 2.5~cm$^{-1}$ spectral resolution.

Fig. \ref{Raman} shows the Raman spectra taken on the Bi$_{2}$Se$_{3}$ flake about 5$\mu$m away from the Pd line (position 1), on top of the Pd line (position 2) and on the Bi$_{2}$Se$_{3}$ flake directly adjacent to the Pd line (positions 3 and 4). In the spectra recorded at positions 1, 3 and 4 we observe phonon bands at 73, 132, and 176~cm$^{-1}$, which is typically found for Bi$_2$Se$_3$\cite{Richter1977}. The widths of the lines are 5 to 8~cm$^{-1}$, which is also typical for bulk Bi$_2$Se$_3$. These results confirm that the crystal structure of the material is unchanged at these positions. We do not observe the Raman bands directly on top of the Pd line (position 2) because the Pd layer on the surface of the Bi$_2$Se$_3$ flake prevents the propagation of the visible excitation light.

The combination of optical, TEM, EDS and Raman spectroscopy with electrical measurements shows that Pd is absorbed by the Bi$_2$Se$_3$ crystal only in the targeted areas, allowing us to pattern superconductivity in Bi$_2$Se$_3$. More work is needed to determine the nature of superconductivity in Pd-doped Bi$_2$Se$_3$. Regardless of whether the patterned regions show topological superconductivity in their own right, or just provide conventional superconductivity in proximity to topological insulator, the patterning provides a promising platform for building novel 2D topological devices.

\section*{Acknowledgments}
%Acknowledgements
This work was supported by the National Science Foundation through grants DGE-1232825 (J.T.M.) and DMR-1507782. We would like to thank the JHU Raman Spectroscopy Users Center.

\end{document}